\documentclass[%
 aip,
rsi,%
 amsmath,amssymb,
 reprint,%
twocolumn
]{revtex4-2}

\usepackage{graphicx}
\usepackage{siunitx}
\DeclareSIUnit{\dbm}{dBm}
\usepackage{upgreek}
\usepackage{xcolor}
\usepackage[utf8]{inputenc}
\usepackage{physics}
\usepackage{hyperref}

\begin{document}

\preprint{AIP/123-QED}

\title{Absorptive filters for quantum circuits: Efficient fabrication and cryogenic power handling}

\author{Alexandre Paquette}
\affiliation{Institut Quantique, Université de Sherbrooke, Sherbrooke, Québec J1K 2R1, Canada}
\author{Joël Griesmar}
\affiliation{Institut Quantique, Université de Sherbrooke, Sherbrooke, Québec J1K 2R1, Canada}
\author{Gabriel Lavoie}
\affiliation{Institut Quantique, Université de Sherbrooke, Sherbrooke, Québec J1K 2R1, Canada}
\author{Romain Albert}
\affiliation{Univ.~Grenoble Alpes, CEA, INAC-PHELIQS, F-38000 Grenoble, France}
\author{Florian Blanchet}
\affiliation{Univ.~Grenoble Alpes, CEA, INAC-PHELIQS, F-38000 Grenoble, France}
\author{Alexander Grimm}
\affiliation{Univ.~Grenoble Alpes, CEA, INAC-PHELIQS, F-38000 Grenoble, France}
\author{Ulrich Martel}
\affiliation{Institut Quantique, Université de Sherbrooke, Sherbrooke, Québec J1K 2R1, Canada}
\author{Max Hofheinz}
\affiliation{Institut Quantique, Université de Sherbrooke, Sherbrooke, Québec J1K 2R1, Canada}
\affiliation{Univ.~Grenoble Alpes, CEA, INAC-PHELIQS, F-38000 Grenoble, France}

\date{\today}

\begin{abstract}

  We present an efficient fabrication method for absorptive microwave filters based on Eccosorb CR-124. Filters are fabricated from readily available parts and their cut-off frequency can be set by their length. They exhibit desirable properties such as a very large and deep stop band with rejection beyond \SI{120}{\decibel} at least up to \SI{40}{\giga\hertz}, more than \SI{10}{\decibel} return loss in both the pass and the stop band as well as an error-function shaped step response without overshoot.
  
  Measurements at very low temperature show that the filters thermalize on a time scale of the order of \SI{100}{\second} and that they can absorb power as high as \SI{100}{\nano\watt} with their noise temperature staying remarkably cool, below \SI{100}{\milli\kelvin}. 

  These properties make the filters ideal for cryogenic filtering and filtering of IF port signals of mixers.
\end{abstract}


\maketitle

\section{Introduction}

Distributed low pass filters based on lossy transmission lines have proven indispensable for filtering lines accessing quantum devices in the microwave regime, such as superconducting quantum circuits or quantum dots. They allow for very deep and wide stop bands, impossible to achieve with lumped elements due to their parasitics, or with non-dissipative distributed circuits due to higher modes of the resonant structures in the filters.

Rejection of these lossy-transmission-line filters is achieved by absorption instead of reflection, allowing for impedance matching not only in the passband but also in the stop band. This is desirable when filters are used next to very nonlinear devices such as circuits containing Josephson junctions. The matching can also help thermalize electromagnetic modes in quantum circuits. The downside is that due to the fluctuation dissipation theorem such filters must emit noise at frequencies below $kT/h$. The temperature of the absorptive medium is, therefore, critical for quantum device performance. 

A popular way of implementing lossy-transmission-line filters is to use the frequency dependence of the surface resistance of the conductors of a transmission line. It increases with frequency due to decrease of the skin depth. A common geometry are metal powder filters where an insulated wire is embedded in a resistive medium formed by metal powder \cite{martinis87, lukashenko08, mueller13, scheller14, lee16, lee18}, but carbon nanotubes have also been used\cite{moghaddam19}.  These filters typically require transmission-line lengths in the meter range for cut-off frequencies in the \SI{100}{\mega\hertz} range. For strong attenuation, the dielectric must be as thin as possible. Impedance matching can be achieved in this type of filter by embedding metal powder \cite{milliken07, wollack14} or other conductive particles \cite{moghaddam19} in an epoxy matrix and carefully adjusting its density.

Alternatively, the surface impedance of the wire itself can be used\cite{vion95,zorin95,lesueur06,spietz06,bluhm08,tancredi14}. In this case, strong attenuation and compact formats can be achieved by using very resistive wires and thin dielectrics\cite{vion95}.

Another method consists of loading the dielectric of a transmission line with magnetic particles\cite{santavicca08, slichter09}. In that case, loss arises dominantly from the hysteresis in the magnetization of the particles. These filters can have much higher loss per unit length, steeper roll-off and typically require lengths of only \SI{0.1}{\meter} to achieve cut-off frequencies in the \SI{100}{\mega\hertz} range. In order to achieving impedance matching with this type of filter, different carefully tuned transmission line geometries have been proposed \cite{santavicca08,slichter09}. 

Some authors have reported that such filters thermalize down to \SI{20}{\milli\kelvin}\cite{slichter09}. However, questions remain as to how well loaded dielectrics can dissipate power at cryogenic temperatures, where heat conductivity is dominated by electrons. Loaded dielectrics, where metal particles do not percolate, are expected to have dramatically lower heat conductivity than percolated or sintered metal powders.

We describe here a convenient method for obtaining impedance matched low-pass filters from readily available standard RG402 and RG405 coax cables and Eccosorb CR-124 iron-loaded epoxy. We then show that these filters stay surprisingly cool even at high powers, with noise temperatures well below \SI{100}{\milli\kelvin} at \SI{100}{\nano\watt}.

\section{Fabrication}

Our filters are lossy coax lines with Eccosorb CR-124 \cite{eccosorbCR} as dielectric, allowing to easily set the cut-off frequency, by simply changing its length. Eccosorb CR are 2-part castable expoxies loaded with iron particles and are responsible for most of the attenuation in the finished filters. Our choice of Eccosorb CR-124, the most heavily loaded in the CR series, allows for convenient filter lengths for the desired cut-off frequencies in the \SI{100}{\mega\hertz} to \SI{1}{\giga\hertz} range.

The outer conductor of the filter is made from RG-402 semi-rigid copper coax cable from which the original dielectric and inner conductor are removed. To do so, the desired length, typically between \SI{50}{\mm} and \SI{100}{\mm}, is dipped in liquid nitrogen. When fully cooled, the PTFE dielectric contracts substantially more than the copper so that in can be pulled out easily  using pliers. As center conductor of the filter, \SI{0.51}{\mm} (AWG 24) stainless steel or brass wire was experimentally determined to offer the best matching in the 0 to \SI{10}{\giga\hertz} range of most interest to us. Note that this is also the diameter of the center conductor of RG-405 coax cable. 
We use Southwest Microwave field replaceable SMA connectors for their modular design which makes it easy to cast the Eccosorb.

\begin{figure}
  \centering
  	\includegraphics[width=\columnwidth]{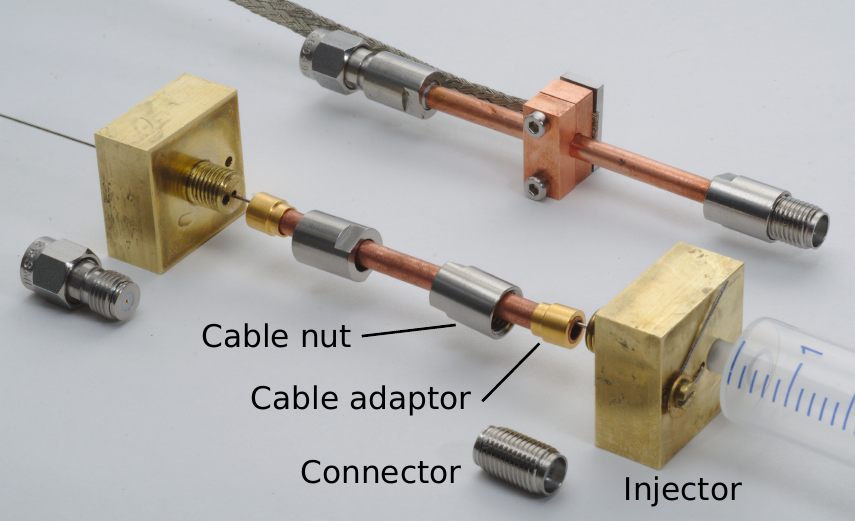}
	\caption{Photo of the \SI{50}{\mm} S.S. filter (top right) and of the assembly setup used to make the filters (bottom left).}
	\label{fig:fab}
\end{figure}

To make the filters, the cable nuts are first slid on the outer conductor and the cable adapters are soldered in place. This allows the cable to be screwed onto custom injectors (see Fig.~\ref{fig:fab}).  
The injectors consist of a block of brass with a male $1/4^{\prime\prime}$-$36$ threaded cylinder matching the cable nuts. The center conductor is inserted through small holes in the center of the cylinder, and screws on the back side of the injector allow tensioning the center conductor to keep it aligned.

Eccosorb is then prepared as per manufacturer recommendation, degassed in a vacuum chamber and heated to around \SI{65}{\celsius} in order to decrease its viscosity. Then, using a syringe, it is injected into the filters through a slightly off-centered hole in one of the injectors. The epoxy is then cured at around \SI{65}{\celsius} for at least 12 hours. Prior to assembly, the injectors are coated with mold release wax to allow the cured filters to easily detach from  the injectors. Field-replaceable connectors are finally screwed onto the cable after cutting the center conductor as per mounting instructions.  
Since the new central conductor is smaller than the usual \SI{0.91}{\mm} diameter of the original conductor of the RG402 cable, we use the male and female connectors for RG405 coax which has a center conductor diameter of \SI{0.51}{\mm}, identical to the new center conductor.

For low temperature measurements, a copper strap is clamped to the center of the filter and heat sunk to the cold plate of the dilution refrigerator.

For the rest of the manuscript, we focus on the analysis of three filters. Two of them are \SI{50}{\mm} in length, with one having a brass central conductor and the other a stainless-steel central conductor. The third filter has a length of \SI{100}{\mm} and has a stainless steel central conductor. The filters will later be referred respectively as \SI{50}{\mm} Brass, \SI{50}{mm} S.S. and \SI{100}{mm}. We find stainless steel easier to use since its greater strength makes it unlikely to break when unmolding.

\section{Room temperature characteristics}

\begin{figure}
	\centering
	\includegraphics[width=\columnwidth]{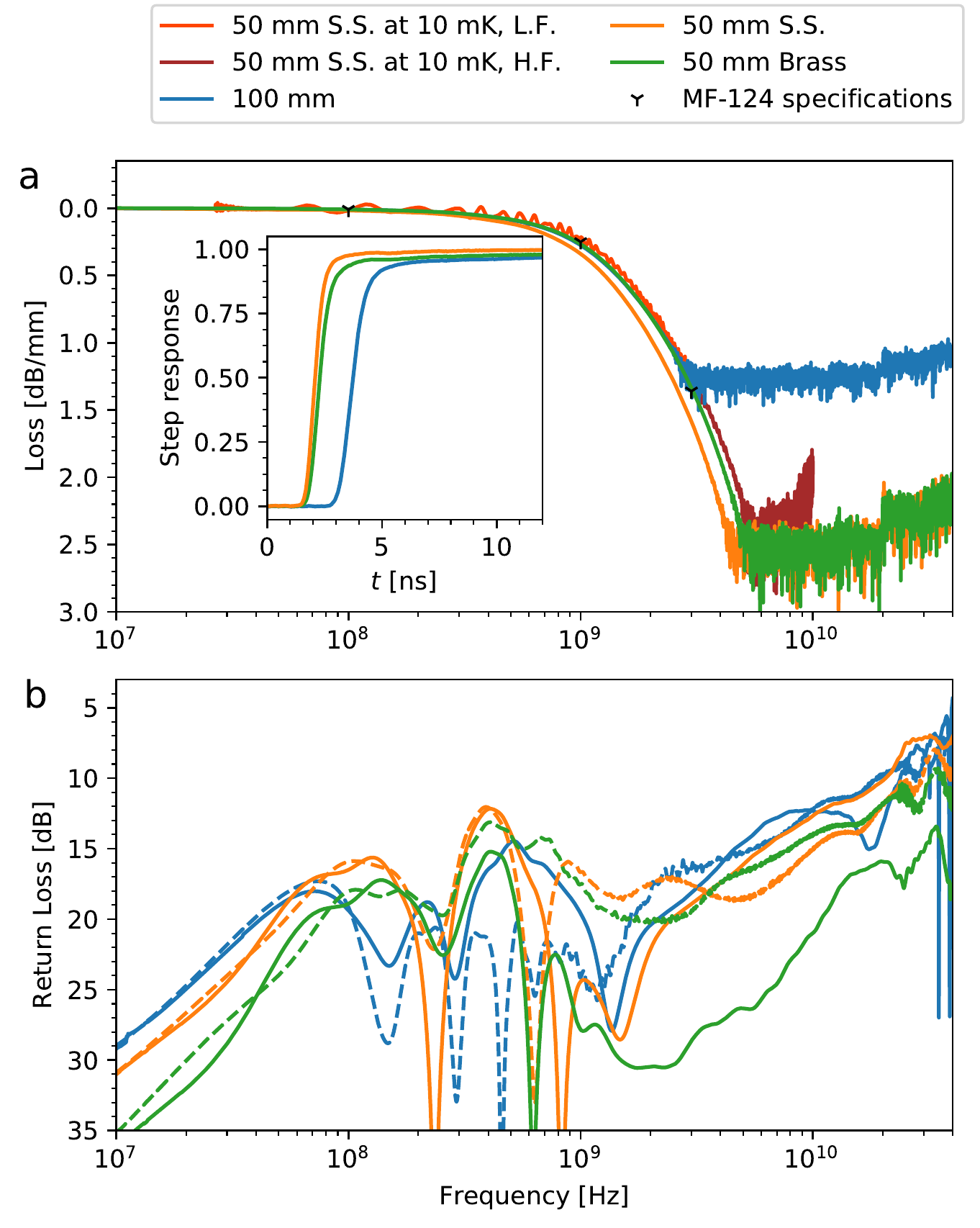}
	\caption{Scattering parameters and step response. All measurements are taken at room temperature except the ones marked \SI{10}{\milli\kelvin}. (a) Insertion loss of the filters measured with a Rohde \& Schwarz ZNB 40 normalized by their length, as well as the attenuation per unit distance specified by the manufacturer\cite{eccosorbMF}. The background noise level of all curves corresponds to approximately \SI{120}{\dB} insertion loss and appears at different loss values depending on filter length. Low temperature measurements use two different ports for low and high frequency measurements (see Fig.~\ref{fig:cryo_setup}) and do not allow for calibration of port match, which explains the ripples in the low-temperature data. The inset shows the step response of the filters, measured by a Tektronix CSA803 with SD-24 sampling head. (b) Return loss of the filters. Solid lines show the male end, dashed lines the female end.}
	\label{Sparam}
\end{figure}

We measure the filter insertion loss and return loss between \SI{10}{\mega\hertz} and \SI{40}{\giga\hertz} at room temperature using a VNA (Rohde \& Schwarz ZNB40).
In order to compare the insertion loss of the filters we have normalized it by their length (see Fig.~\ref{Sparam}a). We observe a very reproducible insertion loss per unit length, meaning that the cut-off frequency of the filter can be simply adjusted by choosing the desired length. The \SI{50}{\mm} S.S, \SI{50}{\mm} Brass and \SI{100}{\mm} filters respectively have a \SI{-3}{\decibel} cut-off frequency of \SI{307}{\mega\hertz}, \SI{395}{\mega\hertz} and \SI{246}{\mega\hertz}. Their loss reaches the VNA dynamic range of approximately \SI{120}{\decibel} between 3 and \SI{5}{\giga\hertz} and rejection stays above \SI{120}{\decibel} at least up to \SI{40}{\giga\hertz}. As seen in the inset of Fig.~\ref{Sparam}a, the filter step response settles quickly and virtually without overshoot. Its error-function shape, makes the filters interesting for filtering base-band pulses, such as Z-gates on superconducting qubits or for driving the IF ports of mixers for X/Y-gates. 

Fig.~\ref{Sparam}b shows that all filters are well impedance matched with return loss above \SI{10}{\decibel} for all filters, or even \SI{15}{\decibel} for the best one, for frequencies from \SI{10}{\mega\hertz} up to \SI{20}{\giga\hertz}. Impedance matching is particularly good in the low \si{\giga\hertz} range which is most relevant for quantum circuits. We attribute slight variations in return loss to imperfect alignment of the center conductor or small voids in the dielectric. 

Above \SI{10}{\giga\hertz}, matching degrades, possibly due to a decrease in the characteristic impedance of Eccosorb CR-124 with frequency \cite{eccosorbMF}. In this range, impedance matching is further complicated by non-TEM modes which we expect to appear above approximately \SI{12}{\giga\hertz}, instead of \SI{34}{\giga\hertz} in RG402, due to the slower propagation speed in Eccosorb with respect to PTFE.

\section{Thermalization at low temperature}

\begin{figure}
	\centering
	\includegraphics[width=\columnwidth]{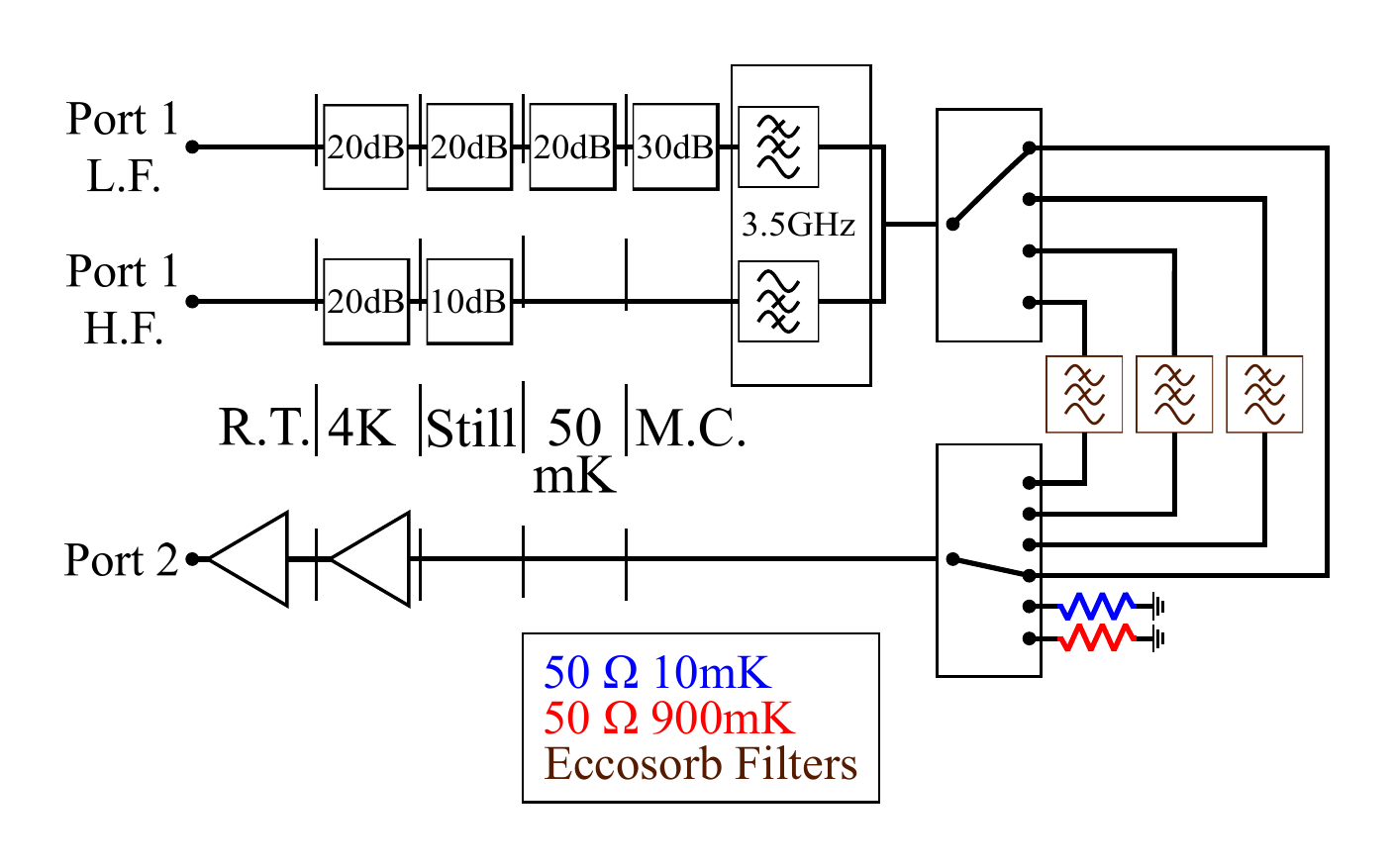}
	\caption{Cryogenic setup.  L.F. and H.F. lines are attenuated and combined at the mixing chamber via a Mini-Circuits ZDSS-3G4G-S+ diplexer with \SI{3.5}{\giga\hertz} cross-over frequency. To allow for measurement of multiple filters in the same cooldown, two Radiall R591\,763\,600 6-ports microwave switches are used. The first amplification stage at \SI{4}{\kelvin} is a Low Noise Factory LNF-LNC0.3\_14A with a gain of approximately \SI{38}{\decibel}.}
	\label{fig:cryo_setup}
\end{figure}

We now address the crucial question of how well these filters thermalize at cryogenic temperatures when subject to noise from room temperature or microwave signals. For these low-temperature measurements we use the setup shown in Fig.~\ref{fig:cryo_setup}. We use two input transmission lines with different attenuation to allow for precise measurements in the pass band as well as in the stop band of the filters. The lines are combined at the base temperature of the dilution refrigerator using a diplexer with a transition frequency of \SI{3.5}{\giga\hertz}. The low-frequency line (Port 1 L.F.\ in Fig.~\ref{fig:cryo_setup}) has a total of \SI{90}{\decibel} of attenuation and is used to measure the filter characteristics in the pass band while adding minimal thermal noise. On the other hand, the high frequency line (Port 1 H.F.\ in Fig.~\ref{fig:cryo_setup}) has only \SI{30}{\decibel} of attenuation and is used to measure the filter characteristics in the stop band. We also use this line to evaluate the power handling of the filters. 
To easily measure all the filters during the same cooldown and to calibrate the setup, we connect them through two 6-port switches. Two of the ports of the switch connecting the filters to the amplifier are terminated with \SI{50}{\ohm} loads. They are thermally isolated via short sections of superconducting NbTi-NbTi coax cable. One of them is then thermalized to the mixing chamber while the other is thermally anchored to the still at \SI{900}{\milli\kelvin}. We use these resistors to calibrate the gain and noise of the amplification chain. The two switches are also connected with a cable the same length and properties as the ones used to connect the filters. It is used to calibrate VNA transmission measurements, as well as the injected power, once the gain of the amplification chain is known.

With the setup calibrated, we first observe that the filter response at low temperature is very close to the one at room temperature, but with slightly higher cutoff freqency of approximately \SI{460}{\mega\hertz} (see Fig.~\ref{Sparam}).
\begin{figure}
	\centering
	\includegraphics[width=\columnwidth]{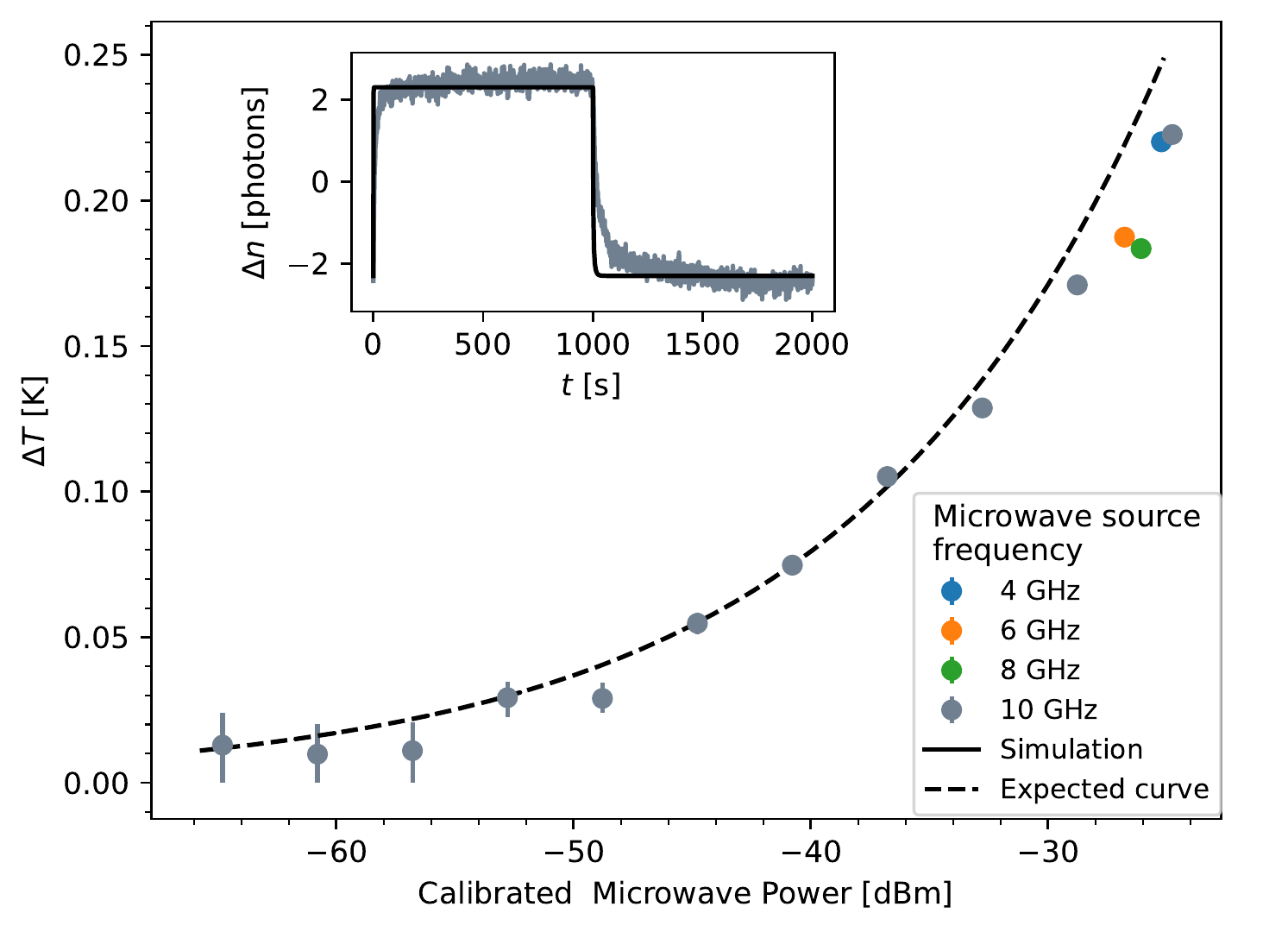}
	\caption{Noise temperature rise $\Delta T$ of the \SI{50}{mm} S.S. filter vs.\ injected microwave power. The expected curve was calculated from \ref{eq:powerhandling}. The inset shows the time evolution of the PSD at \SI{1}{\giga\hertz} with respect to its average as a function of time during one ``on''--``off'' cycle of a microwave tone of \SI{-24.75}{\dbm} at \SI{10}{\giga\hertz}. The simulated response was calculated as described in the S.I. See text for measurement and calibration of injected power.}
	\label{thermalisation}
\end{figure}

We now focus on the power handling of the filters. To do so we inject microwave tones with various amplitudes into the filter. We found it impossible to reliably measure the absolute noise temperature of the filter, because it is much lower than the amplifier noise temperature. To circumvent this problem, we use a differential approach and measure the Power Spectral Density (PSD) between \SI{1}{\giga\hertz} and \SI{5}{\giga\hertz} during \SI{1000}{\second} with a microwave tone and then for the same time and at the same frequencies without a microwave tone.

To help reduce noise, these 2000 seconds ``on''--``off'' cycles were repeated 20 times and then averaged (see inset of Fig.~\ref{thermalisation}). The ``on'' and ``off'' noise levels were then determined by taking the average of the $\operatorname{PSD}(f,t)$ on the last \SI{500}{\second} of each half cycle when the noise temperature has stabilized (see inset of Fig.~\ref{thermalisation}). For our analysis we assume that the photon occupation number is essentially 0 in the ``off'' state, so that the difference between the PSD with and without the microwave tone is directly the number of thermal photons in the ``on'' state. Details on the analysis and a justification of this approximation are given in the S.I.

Fig.~\ref{thermalisation} shows the results of this measurement over a large range of powers with an input tone at \SI{10}{\giga\hertz}. The rise in noise temperature stays below \SI{100}{\milli\kelvin} up to surprisingly high powers of over \SI{100}{\nano\watt}, approximately 3 orders of magnitude higher than the power causing a similar temperature rise in a commercial thin film resistor\cite{huard07}.

We observe that powers below \SI{-55}{\dbm} do not cause a significant temperature rise exceeding \SI{25}{\milli\kelvin}. Note that for these powers we observe a significant number of points where the measured number of photons ($n(f)$) is below 0. We clip these values to $n=0$, resulting in an upward bias on $T$. Therefore, below $\SI{-55}{\dbm}$ all we can say is that we can not detect a significant temperature rise. Because the noise input to the filter is several orders of magnitude lower than $\SI{-55}{\dbm}$ (see SI), we can safely assume that the filters thermalize to $<\SI{25}{\milli\kelvin}$. We also took measurements with a microwave tone at 4, 6 and \SI{8}{\giga\hertz} which agree well with the ones at \SI{10}{\giga\hertz}.

From the inset of Fig.~\ref{thermalisation}, we can also see that the thermalization is far from instantaneous, with the filters' noise temperatures taking about \SI{500}{\second} to fully stabilize with time constants are of the order of \SI{100}{\second} for cooling and \SI{30}{\second} for heating. These long time constants mean that for most pulsed experiments the filter temperature will stay constant and the filter temperature will only depend on the average power. 

In order to better understand the power handling and thermalization dynamics we compare these results with thermal measurements performed by Wikus \emph{et al.} at higher temperature on bulk Eccosorb CR-124 \cite{wikus09}.
By extrapolating their expression for the thermal conductivity of Eccosorb CR-124 to temperatures below \SI{800}{\milli\kelvin} we obtain (see S.I.)
\begin{equation}
  P \approx \lambda_\mathrm{eff} \Delta T^3
  \label{eq:powerhandling}
\end{equation}
with $P$ the injected power and $\lambda_\mathrm{eff} = \SI{2e-4}{\watt\kelvin^{-3}}$ for a \SI{50}{\milli\meter} long filter. This expression is shown as dashed line in Fig.~\ref{thermalisation} and is in excellent agreement with our data without any fitting parameter.

This observed $T^3$ scaling of the power is more favorable than the $T^5$ scaling of electron-phonon coupling in thin film resistors \cite{huard07} and indicates that these filters thermalize significantly better at very low temperatures than attenuators, terminators or other devices based on thin film resistors.

Numerical calculations, depicted in the inset of Fig.~\ref{thermalisation}, extrapolating the expressions for thermal conductivity and heat capacity by Wikus \emph{at al.}~to lower temperatures predict thermalization of approximately \SI{2.3}{\second} for cooling and \SI{0.14}{\second} heating, about 100 times faster than our experimental results (see S.I.). In these calculations we have included the heat capacity of the stainless steel center conductor. The large discrepancy is likely due to a combination of two causes: The filter may heat up the center conductors of the coax cables or other components to which it is connected. In addition, as Wikus \emph{at al.} suspect, Eccosorb may preset a Schottky anomaly, and have a significantly higher heat capacity than the extrapolation of their measurements to lower temperature. We therefore conclude that the thermal conductivity of Eccosorb calculated by Wikus can be extrapolated to dilution refrigerator temperatures, but likely not the heat capacity. 

\section{Conclusion}
In conclusion we have demonstrated a simple fabrication method for impedance-matched dissipative low-pass filters. The filters combine good impedance matching in the pass band as well as the stop band with deep and wide stop band and rapidly settling impulse response. At cryogenic temperatures, our measurements show that the filters have long thermal time constants of the order of \SI{100}{\second} and can handle nanowatt average powers without significant noise temperature rise, so that their noise temperature will remain constant and close to base temperature in typical pulsed experiments.
This makes these filters ideal for a wide range of applications in quantum engineering from IF port filtering of diode mixers at room temperature to filtering flux bias lines of superconducting qubits below \SI{20}{\milli\kelvin}.

\begin{acknowledgments}
  This work was supported by the Natural Sciences and Engineering Research Council of Canada, the Canada First
Research Excellence Fund. 
\end{acknowledgments}

The s-parameter data that supports the findings of this study are available within the article and its supplementary material. Low-temperature data is available upon reasonable request. 

\setcounter{figure}{0}
\renewcommand\thefigure{S\arabic{figure}}
\section{Supplementary Information}

\subsection{Effective filter temperature}

The effective temperature of the mode at frequency $f$ is obtained by inverting the Bose occupation factor:

\begin{equation}
  T(f)=\frac{hf}{k~\ln \left(1+\frac{1}{n(f)}\right)}.
  \label{eqn:Temp}
\end{equation}

\begin{figure}
	\centering
	\includegraphics[width=\columnwidth]{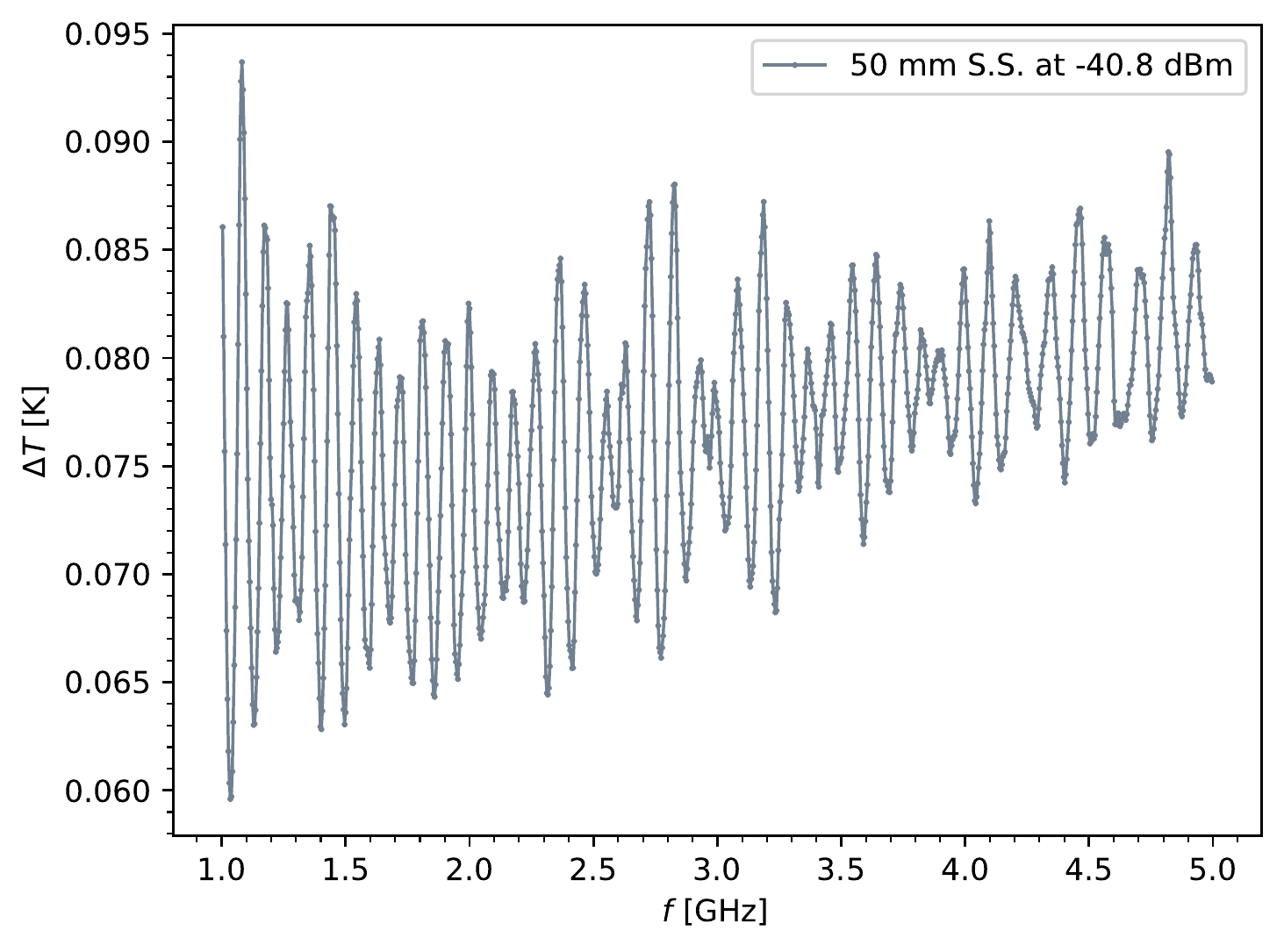}
	\caption{Frequency dependence of the heating of the 50 mm S.S. filter with a \SI{-40.8}{\dbm} microwave tone at \SI{10}{\giga\hertz}.}
	\label{tvsf}
\end{figure}

We find that the noise temperature shows some oscillation as a function of frequency, likely due to standing waves in our setup (see Fig.~\ref{tvsf}), but is globally frequency-independent. 

We then perform a weighted average over different frequencies:
\begin{equation}
	\bar{T}=\sum_f w_f T(f).
\end{equation}

The weights $w_f$ are chosen to minimize statistical uncertainty of the noise temperature:
\begin{equation}
  w_f = \frac{\left(\left.\frac{\partial T}{\partial n}\right|_f\right)^{-2}}{\sum_{\nu} \left(\left.\frac{\partial T}{\partial n}\right|_\nu \right)^{-2}}.
\end{equation}

Due to the amplifier noise we cannot measure $n(f)$ directly, only the change in $n(f)$ between ``on'' and ``off'' states of the microwave tone. In our analysis we therefore have to suppose that $n_\mathrm{off}(f) \approx 0$. With the data of Fig.~\ref{thermalisation} at hand, we can justify this approximation a posteriori.  Neglecting cable attenuation, and non-TEM modes of the coax, the noise power the filter is receiving via the TEM mode of the coax is the cumulative thermal noise from \SI{300}{\kelvin} attenuated by \SI{30}{\decibel}, from \SI{4}{\kelvin} attenuated by \SI{10}{\decibel} and unattenuated from \SI{1}{\kelvin}. This noise amounts to \SI{-90}{\dbm}, several orders of magnitude lower than the powers where we see an increase in noise temperature. We therefore can safely assume that the filter temperature stabilizes below \SI{20}{\milli\kelvin} in the ``off'' state for a base temperature below \SI{10}{\milli\kelvin}, in agreement with Ref.~\cite{slichter09}. This corresponds to a photon occupation number of $n_\mathrm{off} < 0.1$ at \SI{1}{\giga\hertz} and $n_\mathrm{off} < 0.01$ at \SI{2}{\giga\hertz}.

\subsection{Statistical errors}

After averaging over frequency, the dominant statistical errors in these measurements are low-frequency fluctuations in the background noise, arising mostly from gain fluctuations of the amplification chain. In order to estimate this noise we have measured the input-referred noise over a period of 12 hours with the amplification chain connected to the cold \SI{50}{\ohm} load. We have averaged these measurements over \SI{1000}{\second} bins, corresponding to the duration of a half-cycle of our ``on''-``off'' measurement. The error bars in Fig.~\ref{thermalisation} are the \SI{67}{\percent} confidence intervals based on the variance of this reference measurements. Uncertainties are dramatically higher and asymmetric at lower powers because of the strong non-linearity of the Bose occupation factor.

\subsection{Calculation of the power dissipation capabilities of Eccosorb-124}

We want to compare the power handling we have observed with the bulk thermal conductivity of Eccosorb measured by Wikus et al\cite{wikus09} between \SI{800}{\milli\kelvin} and \SI{2.3}{\milli\kelvin}, $\lambda = \lambda_0 T^2$ with $\lambda_0 = \SI{0.0038}{\watt\meter^{-1}\kelvin^{-3}}$. This analysis is complicated by the fact that the power is dissipated unevenly in the filter. It is mostly dissipated close to the central conductor where the field intensities are highest. Along the filter, at low frequencies, close to the cut-off, the power is dissipated in the whole length of the filter. At higher frequencies it is dissipated only in a short section of the filter close to the input port, but the noise of this part of the filter is then attenuated by the reminder of the filter, compensating for the higher local temperature. To get a rough estimate of the power handling expected according to Wikus' formula, we assume that the power is dissipated at the surface of the center conductor and over the whole length of the filter. 

The radial variation of the temperature in the Eccosorb is then given by:

\begin{equation*}
		\dv{T}{r}=-\frac{P}{2\pi r l}\frac{1}{\lambda}\\
\end{equation*}

Where $P$ is the power dissipated, $l$ the length of the filter and $\lambda=\lambda_0T^2$ the thermal conductivity of Eccosorb. Integration results in:
\newcommand\inner{\mathrm{int}}
\newcommand\ext{\mathrm{ext}}

\begin{eqnarray*}
		\int_{T_\inner}^{T_\ext}\lambda_0 T^2 dt&=&-\frac{P}{2\pi l}\int_{r_\inner}^{r_\ext}\frac{\,dr}{r}\\ 
		P&=&\frac{2\pi}{3} \lambda_0 l\frac{T_\inner^3-T_\ext^3}{\ln\left(\frac{r_\ext}{r_\inner}\right)}
\end{eqnarray*}

If $T_\ext \ll T_\inner$ this can be simplified to

\begin{equation*}
  P \approx \lambda_\mathrm{eff} T_\inner^3
\end{equation*}
with
\begin{equation*}
  \lambda_\mathrm{eff} = \frac{2\pi \lambda_0 l}{3\ln\left(\frac{r_\ext}{r_\inner}\right)} = \SI{2e-4}{\watt\kelvin^{-3}}
\end{equation*}
for our filter. This corresponds is eq.~\ref{eq:powerhandling} in our paper.

\subsection{Thermal dynamics}

In order to calculate thermal dynamics of the filter we numerically solve the radial heat diffusion equation

\begin{equation}
  2 \pi r \rho c(T) \frac{\partial T}{\partial t}=\frac{\partial q}{\partial r} = -\frac{d}{dr} \left(2 \pi r \lambda(T) \frac{\partial T}{\partial r}    \right)
\end{equation}

Where $q(r)$ is the radial heat flow per unit length, $\rho = \SI{4600}{\kilogram\per\meter^3}$ the density of Eccosorb and $c(T) = c_e T + c_p T^3$ with $c_e=\SI{0.132}{\joule\per\kilogram\per\square\kelvin}$ and $c_p=\SI{0.0053}{\joule\per\kilogram\per\kelvin\tothe{4}}$ its heat capacity from ref \cite{wikus09}.

As boundary conditions we fix

\begin{equation}
  T(r_\ext) = \SI{10}{\milli\kelvin}
\end{equation}

and
\begin{eqnarray}
  q(r_\inner) &=& - 2 \pi r_\inner \lambda{T} \left.\frac{\partial T}{\partial r}\right|_{r_\inner} \\
  &=& \frac{P(t)}{l} + \pi r_\inner^2 \rho_\mathrm{cc} c_\mathrm{cc}(T) \frac{\partial T(r_\inner)}{\partial t} 
\end{eqnarray}

with $\rho_\mathrm{cc} = \SI{7500}{\kilogram\per\meter^3}$ and $c_\mathrm{cc}(T) = c_{e,\mathrm{cc}} T$ with $c_{e,\mathrm{cc}}=\SI{0.5}{\joule\per\kilogram\per\square\kelvin}$ the expected density and heat capacity of the center conductor.


The result of this numerical calculation is compared to the experimental results in the inset of Fig.~\ref{thermalisation}.

\bibliography{eccosorb}

\end{document}